\documentclass[aps,prl,twocolumn,showpacs,superscriptaddress,amsmath,amssymb]{revtex4}
\usepackage{epic}
\usepackage{curves}
\usepackage{graphicx}

\begin{document}


\title{Rapid onset of collectivity in the vicinity of $^{78}$Ni}


\author{M. Lebois}
\author{D. Verney}
\email[]{verney@ipno.in2p3.fr}
\author{F. Ibrahim}
\author{S. Essabaa}
\author{F. Azaiez}
\author{M. Cheikh Mhamed}
\author{E. Cottereau}
\author{P.V. Cuong}
\altaffiliation{Present address: Institute of Physics, Vietnamese Academy of Sciences and Technology, Hanoi, Vietnam}
\author{M. Ferraton}
\author{K. Flanagan}
\author{S. Franchoo}
\author{D. Guillemaud-Mueller}
\author{F. Hammache}
\author{C. Lau}
\author{F. Le Blanc}
\author{J.-F. Le Du}
\author{J. Libert}
\author{B. Mouginot}
\author{C. Petrache}
\author{B. Roussi\`ere}
\author{L. Sagui}
\author{N. de S\'er\'eville}
\author{I. Stefan}
\author{B. Tastet}
\affiliation{Institut de Physique Nucl\'eaire CNRS-IN2P3/Univ. Paris-Sud 11, Orsay, France}


\date{\today}

\begin{abstract}
$\gamma$-rays following the $\beta$ and $\beta$-n decay of the very neutron rich $^{84}_{31}$Ga$_{53}$ produced by photo-fission of $^{238}$U have been studied at the newly built ISOL facility of IPN Orsay: ALTO. Two activities were observed and assigned to two $\beta$-decaying states: $^{84g}$Ga, $I^\pi=(0^-)$ and $^{84m}$Ga, $I^\pi=(3^-,4^-)$. Excitation energies of the $2^+_1$ and $4^+_1$ excited states of $^{84}_{32}$Ge$_{52}$ were measured at $E(2^+_1)=624.3$ keV and $E(4^+_1)=1670.1$ keV. Comparison with HFB+GCM calculations allows to establish the collective character of this nucleus indicating a substantial N=50 core polarization. The excitation energy of the $1/2^+_1$ state in $^{83}_{32}$Ga$_{51}$ known to carry a large part of the neutron $3s_{1/2}$ strength was measured at 247.8 keV. Altogether these data allow to confirm the new single particle state ordering which appears immediately after the double Z=28 and N=50 shell closure and to designate $^{78}$Ni as a fragile and easily polarized doubly-magic core. 
\end{abstract}


\maketitle

Considerable efforts have been recently deployed in order to reach experimentally the region in the immediate vicinity of $^{78}$Ni to assess the doubly magic character of this very neutron rich nucleus. One of the most interesting questions related to the persistence of the Z=28 and N=50 shell gaps far from stability is whether $^{78}$Ni can be considered as an inert core (a good core) for the nuclear shell model. This is not only crucial for the development of nuclear shell model calculations in this exotic region but also for the underlying physical picture this model carries: Doubly magic nuclei are considered literally as the supporting pillars of the common understanding of nuclear structure. More technically, this question can be put in terms of \textit{polarizability} of the doubly magic core: How fast can the residual interaction between valence nucleons attenuate or even erase the magic character of a core? The most striking example of polarization effects is that of $^{80}$Zr, which in spite of its well known rotational character is effective as a core when a sufficient number of nucleons are present in the valence space (\cite{cau} and Refs. therein). The evidence for a strong Z=28 proton core polarization has been repeatedly pointed out recently in the very neutron rich Ni and Zn isotopes: $^{70}$Ni \cite{per3} and $^{80}$Zn \cite{vdw}. A similar feature for the N=50 neutron core towards $^{78}$Ni has not been observed or discussed with such clarity though first evidence of it can be found in the observation of the mean square charge radius evolutions of $_{38}$Sr \cite{buc}, $_{37}$Rb \cite{thi} and $_{36}$Kr \cite{kei} isotopes. Nuclear structure data for nuclei situated beyond the N=50 shell closure with Z$<$36 is critically required to address this question properly. In this Letter we report on the study of $\gamma$-rays following the $\beta$ and $\beta$-n decays of $^{84}_{31}$Ga$_{53}$ providing the first structure data for the mostly unknown valence space which opens up just above $^{78}$Ni.\par
\begin{figure}
\resizebox{0.5\textwidth}{!}{
\includegraphics*{fig1.eps}
}%
\caption{\label{fig:1}Part (0 keV to 1700 keV) of the $\beta$-gated $\gamma$ spectrum recorded at mass 84. The insert is a zoom of the spectrum for the 600-750 keV energy range. Peaks marked with dots correspond to the decay of $^{90-96}$Rb isotopes stopped within the mass-separator (see text).\\ \\}
\resizebox{0.4\textwidth}{!}{
\includegraphics*{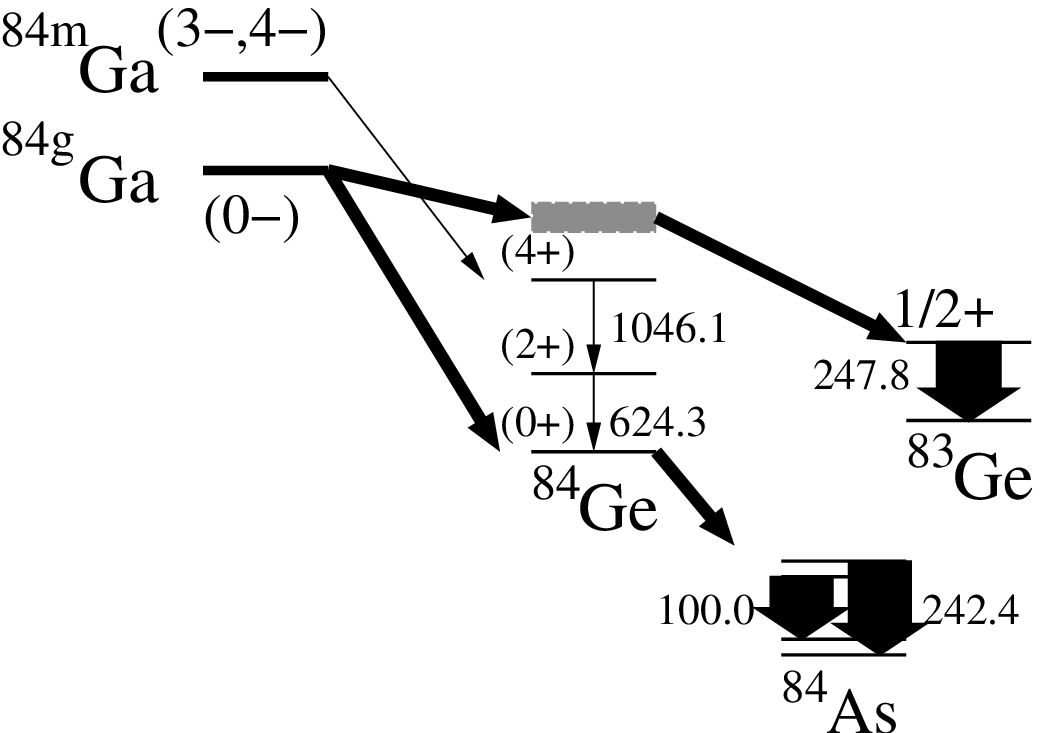}
}%
\caption{\label{fig:chain} Main decay paths of the two identified $\beta$-decaying states. Large arrows indicate the dominant activity in our spectra.}
\end{figure}
$^{84}$Ga isotopes were produced using photo-nuclear reactions at the on-line PARRNe mass-separator operating within the ALTO facility \cite{ibr}. The fission fragments were produced in the interaction of the 50-MeV electron beam delivered by the ALTO linear accelerator with a thick target containing 72 g of uranium in a standard UC$_{x}$ form. The oven was connected to a W ionizing tube heated up to $\simeq2200^\circ$C which selectively ionized alkalies but also elements with particular low first ionization potentials such as Ga and In. The ions were accelerated through 30 kV and magnetically mass separated before being implanted on to a Al-coated mylar tape close to the detection setup. The use of such an ionization mechanism like surface ionization plus mass selection ensured, in the mass region of interest, an A and Z-selective radioactive $1^+$ ion beam production. The suppression of neutron deficient isotopes with photo-fission facilitates the production of a radioactive beam without isobaric contamination allowing sensitive measurements to be made far from stability. Even though a W tube was used, with a modest ionization efficiency for Ga of $0.7\%$, it is still possible to study $^{84}$Ga due to the absence of $^{84}$Rb. These measurements were performed during a test run dedicated to safety measurements scheduled in the different phases of the commissioning of ALTO. For this reason the electron beam intensity was limited to just 1 $\mu$A instead of the nominal 10 $\mu$A. At this primary intensity, the fission rate inside the target is estimated to be $10^{10}$ /s \cite{ibr} and the yield of $^{84}$Ga ions measured at the tape station was of the order of a few tens per second and in agreement with the predicted yields \cite{leb}. The $\gamma$-detection system consisted of one coaxial large volume HPGe detector ($70\%$ relative efficiency) with a resolution of 2.3 keV at 1.3 MeV and one small EXOGAM CLOVER detector \cite{aza} from the prototype series ($100\%$ relative efficiency) with a typical resolution for the central signal of a single crystal of 2.0 keV at 1.3 MeV. These two detectors were placed in a $180^\circ$ geometry close to collection point onto the tape. In this configuration a total photopeak efficiency of $3\%$ at 1.3 MeV was achieved. This collection point was surrounded by a tube-shaped plastic scintillator for $\beta$ detection ($\epsilon_\beta\simeq 30 \mbox{-} 50\%$). In an experiment with an isobaric-pure beam, the tape cycling is used to help in the identification of unknown new lines which are expected to appear in the spectra by analyzing their time dependence. The tape sequence was formed by a 9 s build-up phase followed by a 1 s decay time, evacuation of the source and repetition \textit{ad libitum}.\\
The $\beta$-gated $\gamma$-spectrum which was recorded with the separator field set to A=84 is shown in Fig. \ref{fig:1}. $\gamma$-lines belonging to the decay of neutron rich Rb isotopes with mass ranging from 90 to 96 have been identified in the spectra (dots in Fig. \ref{fig:1}). The Rb isotopes which are strongly favoured in both the fission process and the ionization mechanism ($\epsilon_i\simeq80\%$) were stopped within the separator chamber which was intentionally not shielded to allow the simultaneous safety measurements with the required conditions. The activity of the Rb isotopes was such that the associated $\gamma$-lines could appear as random coincidences with $\beta$ events. Despite this difficult experimental environment, the $\gamma$-lines characterizing the activity of the $\beta$ daughters : $^{84}$Ge (T$_{1/2}=$ 947(11) ms) and $^{84}$As (T$_{1/2}=$ 4.5(2) s) are clearly visible establishing that $^{84}$Ga was successfully collected onto the tape. We confirm the existence of two $\gamma$-rays in the decay of $^{84}$Ga, one at 624.3(7) keV and the other at 1046.1(7) keV (see Fig. \ref{fig:1}), already observed at ISOLDE having a fast decaying component \cite{kos}. Spectra obtained at ISOLDE at A=85 show also the presence of the 624.3 keV peak coming from the $\beta$-n decay of $^{85}$Ga. Then the 624.3 keV peak necessarily corresponds to a transition in $^{84}$Ge and should correspond to the $2_1^+\rightarrow0_{gs}^+$ transition. The $\beta$-n channel is known to be widely open in the decay of $^{84}$Ga with a probability of $P_n=70\pm15\%$ \cite{kra} then $\gamma$-rays from transitions in $^{83}$Ge and its A=83 daughters must be present in the spectra. The 306.3 keV line from the decay of $^{83}$Ge ($I_{\gamma}=100$ relative intensity) and the 1238.2 keV line already attributed to the $\beta$-decay of $^{83}$Ga \cite{per} are indeed visible (see Fig. \ref{fig:1}). It is expected that a peak should appear close to the energy of the $1/2_1^+$ excited state observed in $^{82}$Ge(d,p)$^{83}$Ge direct reaction at $E_x=280\pm20$ keV \cite{tho}. There exists a peak at 247.8(3) keV clearly visible in our spectra and previously not reported in decays of neutron rich A=83 or A=84 isobars which is a good candidate. In the ISOLDE data this peak was very close to the 248.0 keV line due to the decay of the neutron deficient $^{84}$Rb isomeric state ($T_{1/2}=20.26$ m) thus contaminated. Such isobaric contamination is not possible at ALTO due to the photo-fission based production mechanism. The observed 247.8 keV-peak in this data set was determined to have a half-life of $T_{1/2}=75\pm33$ ms, which is compatible with the previously measured $^{84}$Ga half-life $T_{1/2}=85\pm10$ ms \cite{kra}. We propose then to attribute it to the transition $1/2_1^+\rightarrow5/2_{gs}^+$ in $^{83}$Ge which could not be observed in our previous $^{83}$Ga $\beta$-decay experiment due to the very low expected feeding from the supposed $I^\pi=5/2^+$ $^{83}$Ga ground state \cite{per}. The two $\gamma$-lines at 100.0 keV and 242.4 keV attributed to the decay of $^{84}$Ge \cite{omt} are also clearly observed in our spectra. They correspond to transitions between states of $^{84}$As as well as two others (previously not reported in literature) at 42.7(3) keV and 386.0(5) keV.\par
Assuming a unique origin \textit{i.e.} a unique $\beta$-decaying state in $^{84}$Ga which is simultaneously responsible for (\textit{i}) the feeding of the transition $2_1^+\rightarrow0_{gs}^+$ at 624.3 keV in $^{84}$Ge (\textit{ii}) the creation of a population of $^{84}$Ge which in turn decays through the 100.0 keV and 242.4 keV transitions in $^{84}$As and (\textit{iii}) the feeding of the $1/2_1^+\rightarrow5/2_{gs}^+$ in $^{83}$Ge, a clear unbalance in the $\gamma$-activities of the different members of the A=83,84 isobaric chains by a factor 20 is observed. However, the 100.0 keV and 242.4 keV-peak surfaces are quite compatible with that of the 247.8 keV-peak fed in the $\beta$-n channel if one considers the known $^{84}$Ga $P_n$ value of $70\pm15\%$ \cite{kra}. This leads to the conclusion that the largest part of the activity observed in this experiment corresponds to the decay of a low spin state which feeds directly the $^{84}$Ge ground state through the $\beta$-channel and the $^{83}$Ge $1/2_1^+$ state through the $\beta$-n-channel. It should be noted that the activity balance is quite different in the data from ISOLDE where the 624.3 keV and the 247.8 keV transitions were fed equally \cite{kos}. Furthermore, in the ISOLDE A=84 spectra, one clearly observes the feeding of the 866 keV transitions which was already attributed to the $(7/2_1^+)\rightarrow5/2_{gs}^+$ transitions in $^{83}$Ge \cite{per}. Such feeding can only originate from the $\beta$-n-decay of a state with $I\ge 3$ in $^{84}$Ga. From this cross comparison between measurements made in experiments at ISOLDE and ALTO, the existence of \textit{two} $\beta$ decaying states in $^{84}$Ga is strongly favoured: One with very low spin, dominant at ALTO, and a second, well produced at ISOLDE, of spin $I\ge 3$. This explanation, represented in a pictorial way in Fig. \ref{fig:chain}, accounts for the whole experimental data in a coherent way. Now one must come back to the case of the 1046.1 keV peak which is observed at mass 84 with the same intensity as the 624.3 keV peak, but not observed at mass 85 nor at mass 83. If this peak corresponded to a transition in $^{83}$Ge fed by the $\beta$-n decay of the $^{84}$Ga low spin state, then, considering that it is not observed directly in the $^{83}$Ga decay in the experiment of Ref. \cite{per}, it would necessarily be the $1/2_2^+$ state and it should have a similar intensity as the 247.8 keV $\beta$-n peak in the A=84 spectra in the present experiment \textemdash which is not the case. If this 1046.1 keV peak was fed by the $^{84}$Ga high spin state decay, then, considering that it is not observed directly in the $^{83}$Ga decay in the experiment of Ref. \cite{per}, it would necessarily be the $9/2_1^+$ state and it would be connected to the $7/2_1^+$ level at 866 keV by a 799 keV transition that would have been seen in the ISOLDE experiment \textemdash which is also not the case. Therefore the 1046.1 keV-peak belongs to the $^{84}$Ge level scheme. Furthermore, since this 1046.1 keV peak is observed at mass 84 and not mass 85, the only remaining possibility is that it corresponds to the $\gamma$-decay of a $^{84}$Ge excited state of spin higher than 2 (which in turn cannot be fed in the $^{85}$Ga $\beta$-n decay). Since it has the same relative intensity as the 624.3 keV peak, we propose to attribute this 1046.1 keV peak to $4_1^+\rightarrow2_{1}^+$ transition in $^{84}$Ge.\par
\begin{figure}
\resizebox{0.5\textwidth}{!}{
\includegraphics*{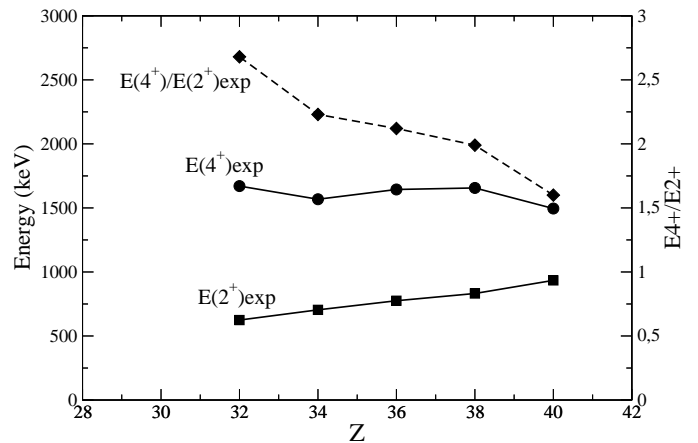}
}%
\caption{\label{Fig:4sur2} Experimental systematics of the energies of the first $2^+$ and $4^+$ states and ratio E($4^+_1$)/E($2^+_1$) for even-even N=52 isotones.}
\end{figure}
\begin{figure}
\resizebox{0.5\textwidth}{!}{
\includegraphics*{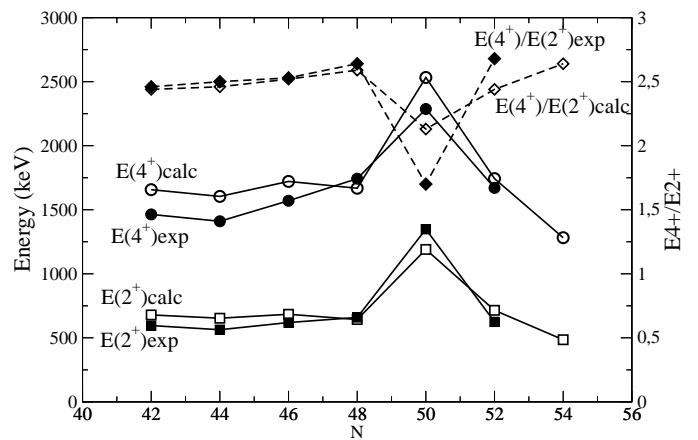}
}%
\caption{\label{fig:syst} Calculated (open symbols) \textit{vs} experimental (filled symbols) systematics of the energies of the first $2^+$ and $4^+$ states and ratio E($4^+_1$)/E($2^+_1$) for the neutron rich even-even Ge isotopes up to mass 86.}
\end{figure}
The coexistence of two close lying levels with spin difference $\Delta I\gtrsim 3$ in $^{84}$Ga is easily accounted for by considering the natural valence space available for 3 neutrons and 3 protons above a $^{78}$Ni core. The recent experimental breakthrough in close vicinity of $^{78}$Ni \cite{per,tho,ver} provides indeed a rather clear picture of the single particle level ordering in this very neutron rich region: For protons, $1f_{5/2}$ and $2p_{3/2}$ are close to each other and in this order in energy while for neutrons, $2d_{5/2}$ is the lowest closely followed by $3s_{1/2}$. Then the natural proton-neutron configuration for $^{84}$Ga ground state is $(\pi 1f_{5/2})^3\otimes (\nu2d_{5/2})^3$. Use of the Paar model describing the proton and neutron coupling to the core vibrations \cite{paa} shows that $I^\pi=0^-$ is the favored member of the proton-neutron multiplet which fits well with our experimental findings. The first excited configuration is either $(\pi 1f_{5/2})^3\otimes (\nu2d_{5/2})^2 (\nu 3s_{1/2})^1$ or $(\pi 1f_{5/2})^2(\pi 2p_{3/2})^1\otimes (\nu2d_{5/2})^3$. The first can only produce a doublet $2^-$-$3^-$ unlikely to give rise to an isomeric state while the second configuration produces a state $I^\pi=4^-$ which happens also to be energetically favoured in Paar's model. It seems then natural to attribute it to the isomeric state. These attributions of spin and configurations, and the adequacy of the particle-core vibration coupling picture follow and complete the general trends of the nuclear structure emerging in the region close to $^{78}$Ni in terms of dominant single particle states and dynamics. This latter aspect is of particular interest here. As can be seen from Fig. \ref{Fig:4sur2} there is a continuous increase of the experimental E$(4^+_1)$/E$(2^+_1)$ values for the N=52 isotones from stability towards neutron excess. The inclusion of the new value at Z=32 obtained in our experiments reveals a sudden acceleration in this tendency which should, most generally speaking, be interpreted as a sudden increase of collectivity. If one now compares to the same systematics for the isotopic chain of Ge (filled diamonds on Fig. \ref{fig:syst}) one sees that the E$(4^+_1)$/E$(2^+_1)$ ratio regains approximately the same value in $^{84}$Ge after the N=50 shell closure as it had in $^{80}$Ge. Everything happens as if the N=50 shell closure offered nothing but a local parenthesis of rigidity. 
\begin{figure}
\resizebox{0.5\textwidth}{!}{
\includegraphics*{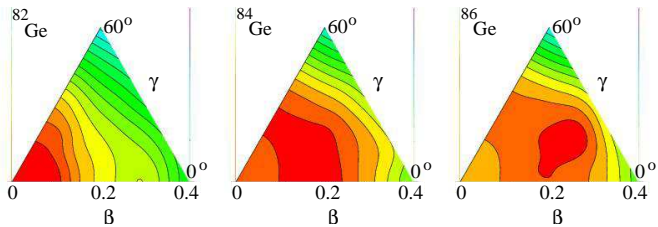}
}%
\caption{\label{fig:pot} Potential energy surfaces $V(\beta,\gamma)$ calculated for for $^{82,84,86}$Ge$_{50,52,54}$ obtained in constrained HFB calculations \cite{per2}. Equipotential lines are separated by 1 MeV.}
\end{figure}\par
In order to understand better the nature of collectivity and its quick onset in this mass region Hartree-Fock-Bogoliubov (HFB) calculations were performed \cite{per2} using the Gogny-D1S effective interaction \cite{dec,ber}. Numerical methods and codes used were those described in Ref. \cite{gir}. Rotational and vibrational degrees of freedom in the triaxial plane were described within the beyond mean-field approach developed by the group of CEA Bruy\`eres-le-Ch\^atel and based on the generator coordinate method (GCM) \cite{lib}. As can be seen in Fig. \ref{fig:syst}, the resulting energy values for the first $2^+$ and $4^+$ excited states are in very good agreement with the experimental ones, as are the ratios E$(4^+_1)$/E$(2^+_1)$. It is striking to consider that even the semi-magic nucleus  $^{82}$Ge is well reproduced in this approach, meaning that the collective vector basis in the GCM calculations is well suited, in other words that quadrupole coherence dominates its structure. The corresponding calculated potential energy maps in the triaxial plane are represented for $^{82-86}$Ge in Fig. \ref{fig:pot} where one clearly sees that the evolution from N=50 to 52 corresponds to a transition from vibration around a spherical equilibrium shape to a complete $\gamma$-softness with the apparition of a marked $\gamma$-valley. Prediction for N=54 is that collectivity increases even more with a surface potential energy closer to that of a triaxial rotor.\par
In conclusion, $^{84}$Ga isotopes were produced on-line with the ALTO facility. We could establish for the first time the existence of a $(3^-,4^-)$ isomeric state in this nucleus which can only originate from the specific single particle relative ordering (quite different from that long known close to stability) appearing in the immediate vicinity of $^{78}$Ni. We observed the consecutive $\beta$-decay of the isomeric state allowing the determination of the energies of the $2^+_1$ and $4^+_1$ excited state of $^{84}$Ge and hinting for an extremely rapid onset of collectivity just after the N=50 shell closure for Z=32. Up to now, the whole corpus of experimental data points towards a doubly-magic $^{78}$Ni with very special \textquotedblleft doubly-polarized\textquotedblright core character.\\
\\
We wish to thank Dr U. K\"oster for fruitful discussions.


\begin{thebibliography}(
\bibitem{cau} E. Caurier \textit{et al.} Rev. Mod. Phys. \textbf{77}, 427 (2005).
\bibitem{per3} O. Perru \textit{et al.}, Phys. Rev. Lett. \textbf{96}, 232501 (2006).
\bibitem{vdw} J. Van de Walle \textit{et al.}, Phys. Rev. Lett. \textbf{99}, 142501 (2007).
\bibitem{buc} F. Buchinger \textit{et al.}, Phys. Rev. C \textbf{41}, 2883 (1990).
\bibitem{thi} C. Thibault \textit{et al.}, Phys. Rev. C \textbf{23}, 2720 (1981).
\bibitem{kei} M. Keim \textit{et al.}, Nucl. Phys. \textbf{A586}, 219 (1995).
\bibitem{ibr} F. Ibrahim \textit{et al.}, Nucl. Phys. \textbf{A787}, 110c (2007).
\bibitem{leb} M. Lebois, Thesis, Univ. Paris-Sud 11 (2008).
\bibitem{aza} F.Azaiez and W.Korten, Nucl. Phys. News \textbf{7}, 21 (1997).
\bibitem{kos} U. K\"oster, communication at the Workshop on Spectroscopy of Neutron-rich Nuclei, Chamrousse, France, 16-20 March 2008.
\bibitem{kra}K.-L. Kratz \textit{et al.}, Z. Phys. A \textbf{340}, 419 (1991).
\bibitem{per} O. Perru \textit{et al.}, Eur. Phys. J. A \textbf{28}, 307 (2006).
\bibitem{tho} J. S. Thomas \textit{et al.}, Phys. Rev. C \textbf{76}, 044302 (2007).
\bibitem{omt} J.P. Omtvedt \textit{et al.}, Z. Phys. A \textbf{338}, 241 (1991).
\bibitem{ver} D. Verney \textit{et al.}, Phys. Rev. C \textbf{76}, 054312 (2007).
\bibitem{paa} V. Paar, Nucl. Phys. \textbf{A331}, 16 (1979).
\bibitem{per2} O. Perru, Thesis, Univ. Paris-Sud 11, IPNO-T05-02 (2004).
\bibitem{dec} J. Decharg\'e and D. Gogny, Phys. Rev. C \textbf{21}, 1568 (1980).
\bibitem{ber} J.F. Berger, M. Girod, and D. Gogny, Comput. Phys. Commun. \textbf{63},365 (1991).
\bibitem{gir} M. Girod, B. Grammaticos, Phys. Rev. C 27 (1983) 2317.
\bibitem{lib} J. Libert, M. Girod, J.-P. Delaroche, Phys. Rev. C 60 (1999) 054301.
\end{thebibliography}
\end{document}